\documentclass[fleqn,10pt]{wlscirep}
\usepackage[utf8]{inputenc}
\usepackage[T1]{fontenc}
\usepackage{subcaption}
\usepackage{array}
\usepackage{ragged2e}
\usepackage{multirow}
\newcolumntype{P}[1]{>{\RaggedRight\hspace{0pt}}p{#1}}

\title{VinDr-Mammo: A large-scale benchmark dataset for computer-aided diagnosis in full-field digital mammography}

\author[1,$\dag$]{Hieu T. Nguyen}
\author[1,2,$\dag$]{Ha Q. Nguyen}
\author[1,2,3,$\dag$,*]{Hieu H. Pham}
\author[4]{Khanh Lam}
\author[5]{Linh T. Le}
\author[1]{Minh Dao}
\author[1,6]{Van Vu}
\affil[1]{Institute of Big Data, Hanoi, Vietnam}
\affil[2]{College of Engineering
and Computer Science (CECS), VinUniversity, Hanoi, Vietnam}
\affil[3]{VinUni-Illinois Smart Health Center,  Hanoi, Vietnam}
\affil[4]{Hospital 108, Department of Radiology, Hanoi, Vietnam}
\affil[5]{Hanoi Medical University Hospital, Department of Radiology, Hanoi, Vietnam}
\affil[6]{Yale University, Department of Mathematics, New Heaven, CT 06511, U.S.A.}
\affil[*]{corresponding author: \textcolor{blue}{hieu.ph@vinuni.edu.vn} (Hieu H. Pham)}
\affil[$\dag$]{these authors contributed equally to this work}

\begin{abstract}
 
Mammography, or breast X-ray imaging, is the most widely used imaging modality to detect cancer and other breast diseases. Recent studies have shown that deep learning-based computer-assisted detection and diagnosis (CADe/x) tools have been developed to support physicians and improve the accuracy of interpreting mammography. 
A number of large-scale mammography datasets from different populations with various associated annotations and clinical data have been introduced to study the potential of learning-based methods in the field of breast radiology. With the aim to develop more robust and more interpretable support systems in breast imaging, we introduce VinDr-Mammo, a Vietnamese dataset of digital mammography with breast-level assessment and extensive lesion-level annotations, enhancing the diversity of the publicly available mammography data.
The dataset consists of 5,000 mammography exams, each of which has four standard views and is double read with disagreement (if any) being resolved by arbitration. The purpose of this dataset is to assess Breast Imaging Reporting and Data System (BI-RADS) and breast density at the individual breast level. In addition, the dataset also provides the category, location, and BI-RADS assessment of non-benign findings. We make VinDr-Mammo publicly available 
as a new imaging resource to promote advances in developing CADe/x tools for mammography interpretation.

\end{abstract}
\begin{document}

\flushbottom
\maketitle

\thispagestyle{empty}

\section*{Background \& Summary} \label{sec:introduction}



Breast cancer is among the most prevalent cancers and accounts for the largest portion of cancer deaths, with an estimated 2.2 million new cases in 2020~\cite{sung2021global}. Treatment is most successful when breast cancer is at its early stage. Biennial screening can reduce breast cancer mortality rate by 30\%~\cite{mandelblatt2016collaborative}. Among standard imaging examinations for breast cancer diagnosis, namely mammography, ultrasound, digital breast tomosynthesis, and magnetic resonance, mammography is the recommended modality for cancer screening~\cite{siu2016screening}. Interpreting mammography for breast cancer screening is a challenging task. The recall rate of mammogram screening is around 11\% with a sensitivity of 86.9\%, while the cancer detection rate is 5.1 per 1,000 screens ~\cite{lehman2017national}. It means that a large portion of cases called back for further examinations eventually result in non-cancer. Improving cancer screening results may help reduce the cost of follow-up examinations and unnecessary mental burdens on patients.

With recent advancements of learning-based algorithms for image analysis~\cite{krizhevsky2012imagenet,lecun2015deep}, several works have adapted deep learning networks for mammography interpretation and showed potential to use in clinical practices~\cite{dembrower2020effect,rodriguez2019stand,rodriguez2019detection,nanwu2020s,dream2020,mckinney2020international}. In retrospective settings, the CAD tool as an independent reader can achieve a performance comparable to an average mammographer~\cite{rodriguez2019stand}. It can be leveraged as a decision support tool that helps enhance radiologists' cancer detection with the reading time being unchanged~\cite{rodriguez2019detection}. In another human-machine hybrid setting, where radiologists and machine-learning algorithm independently estimate the malignancy of the lesions, the linear combination of human and machine prediction show higher performance than a single human or machine reader~\cite{nanwu2020s}. The improvement as a result human-machine combination is also witnessed in screening mammography interpretation~\cite{dream2020}. Furthermore, there was evidence that shows a machine learning model developed by training on data from a specific population (UK) can generalize and perform well on another population (US)~\cite{mckinney2020international}.

The recent progress in the study of mammography interpretation has drawn much attention with an increasing number of mammogram datasets with various characteristics, while some datasets are publicly available to the research community, some have restricted access or are not open \cite{halling2020optimam,dembrower2020multi,chinese,wu2019nyu,moreira2012inbreast,bowyer1996digital,suckling1994mammographic}. Digital Database for Screening Mammography (DDSM)~\cite{bowyer1996digital} and Mammographic Image Analysis Society (MIAS) dataset~\cite{suckling1994mammographic}
are the two earliest public datasets that provide digitalized scans of screen-film mammograms with precise annotations of breast abnormalities. The MIAS dataset was released in 1994 with 161 studies collected in the United Kingdom while the DDSM dataset consisted of 2,620 exams collected from institutions in the United States. Compared to the former one, the DDSM dataset has a significantly larger scale and follows the BI-RADS standard. To the best of our knowledge, INbreast~\cite{moreira2012inbreast}, released in 2012 with 115 exams from Portugal, is the very first public dataset that provides digital mammograms with lesions annotations and overall exam assessment following the BI-RADS standard. In 2019, the NYU Breast Cancer Screening Dataset~\cite{wu2019nyu} was introduced with 229,426 screening exams, consisting of 1,001,093 images, from 141,473 women screened at NYU Langone Health. The dataset contains breast-level cancer based on biopsy results, exam-level assessment of BI-RADS, breast density, and biopsied finding annotations. While the dataset is not public, the subsequent work~\cite{nanwu2020s} based on this dataset showed evidence that a large-scale dataset of mammography can enable a computer-aided system that helps improve radiologist performance. At around the same time, the Cohort of Screen-Aged Women Case-Control (CSAW-CC)~\cite{Strand2022CSAW-CC} was opened for evaluating AI tools for breast cancer, including 1303 cancer cases and 10,000 randomly selected controls from Karolinska University Hospital. The CSAW-CC dataset is a subset of the full CSAW dataset including women screened in the Stockholm region between 2008 and 2015. In cancer cases, visible tumors in mammography were manually annotated on a pixel level. Another large-scale dataset is the OPTIMAM mammography image database~\cite{halling2020optimam} (OMI-DB) which consists of images and clinical data of 172 282 women screened and diagnosed in several institutions in the United Kingdom since 2011. To access to the OMI-DB dataset, the research group must submit an application to elaborate the scientific purpose based on the dataset which will be reviewed by the OPTINAM steering committee. In addition, the Chinese Mammography Database was recently introduced, containing 1,775 studies from several Chinese institutions. All cases have breast-level benign and malignant confirmed by biopsy, and molecular subtypes are available for 749 cases.
A summary of the characteristics of these datasets is given in Table \ref{tab:existing-data}.

\begin{table}[ht]
\footnotesize
\caption{\textsf{Summary of mammography datasets.}}

\setlength{\tabcolsep}{5pt}
\renewcommand{\arraystretch}{1.5}
\begin{tabular}{P{35pt}P{35pt}P{30pt}P{33pt}P{36pt}P{60pt}P{65pt}P{40pt}P{30pt}P{35pt}}
\hline
\textbf{Dataset} & Origin & Introduction year & \#studies & \#images & Finding type & Annotations & BI-RADS assessment & Breast density & Mode of acquisition \\ \hline

\textbf{MIAS~\cite{suckling1994mammographic}} & United Kingdom & 1994 & 161 & 322 & Mass, calcification, asymmetry, and distortion & Circle around the finding, specified by center and radius & No & Yes & SFM\\

\textbf{DDSM~\cite{bowyer1996digital}} & United States & 1999 & 2,620 & 10,480 & Mass and Calcification & Contour enclosing the finding & Yes & Yes & SFM \\

\textbf{INBreast~\cite{moreira2012inbreast}}  & Portugal & 2012 & 115 & 410 & Mass, calcification, asymmetry, and distortion & Contour enclosing the finding & Yes & Yes & FFDM \\

\textbf{NYU Dataset~\cite{wu2019nyu}} & United State & 2019$^{\dagger}$ & 229,426 & 1,001,093 & Biopsied lesions & Contour enclosing the finding & Yes & Yes & FFDM \\

\textbf{CSAW-CC~\cite{dembrower2020multi}} & Sweden & 2020 & 24,694 & 98,788 & Visible tumors \& tumor signs & Contour enclosing the finding & No & No & FFDM \\

\textbf{OMI-DB~\cite{halling2020optimam}} & United Kingdom & 2021 & NA & 3,072,878$^{*}$ & Lesions & Rectangular region of interest  & No & No & FFDM \\

\textbf{CMMD~\cite{chinese}} & China & 2021 & 1,775 & 5,202 & Biopsied abnomalities(mass or calcification & No local annotations & No & No & FFDM \\

\textbf{VinDr-Mammo} & Vietnam & 2022 & 5,000 & 20,000 & Mass, calcification, asymmetry, distortion, and other associated features & Rectangular region of interest & Yes & Yes & FFDM \\

 \hline

\end{tabular}

\textsf{$\dagger$ Not publicly accessible \\ $*$ Including for-presentation and for-processing images}
\label{tab:existing-data}
\end{table}

Along with the existing mammography datasets, we introduce and release the VinDr-Mammo dataset, an open-access large-scale Vietnamese dataset of full-field digital mammography consisting of 5,000 four-view exams with breast-level assessment and extensive lesion-level annotations. Our aims is to enhance the diversity of the publicly available mammography data for a more robust AI system and to lean towards a more interpretable system via extensive lesion-level annotations. Mammographies were acquired retrospectively from two primary hospitals in Hanoi, Vietnam, namely Hospital 108 (H108) and Hanoi Medical University Hospital (HMUH). Breast cancer assessment and density are reported following Breast Imaging Reporting and Data System~\cite{birads_2013}. Breast abnormalities that need short-term follow-up or are suspicious of malignancy are marked by bounding rectangles. Following European guideline~\cite{amendoeira2013european}, mammography exams were independently double-read. Any discordance between the two radiologists would be resolved by arbitration with the involvement of a third radiologist. To the best of our knowledge, VinDr-Mammo is currently the largest public dataset (20,000 scans) of full-field digital mammography that provides breast-level BI-RADS assessment category along with suspicious or probably benign findings that need follow-up examination. By introducing the dataset, we contribute a benchmarking imaging dataset to evaluate and compare algorithmic support systems for breast cancer screening based on FFDM.

\section*{Methods}

\begin{figure}[h]
    \centering
    \includegraphics[width=0.9\linewidth]{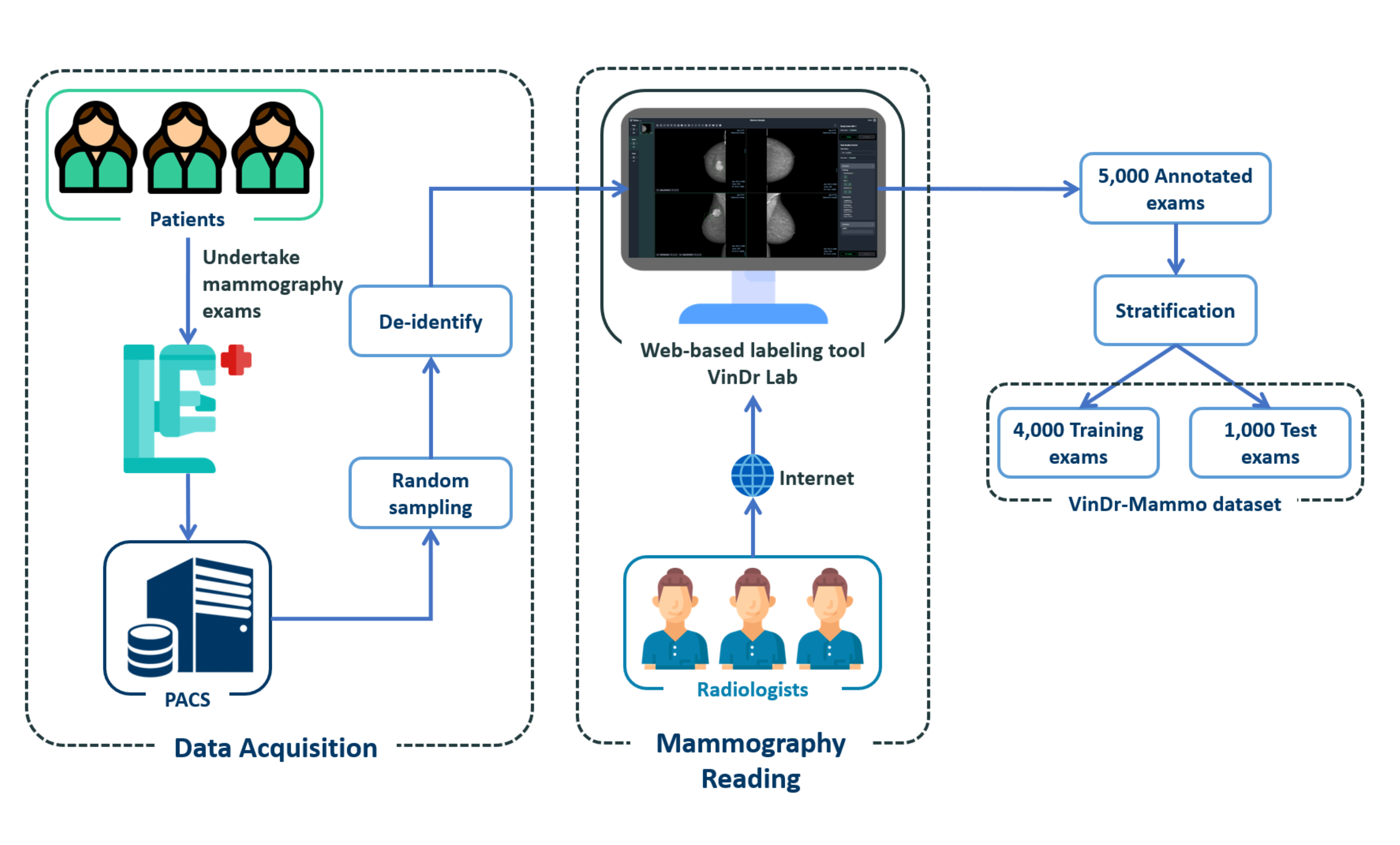}
    \caption{Overview of the data creation process. First, for-presentation mammograms in DICOM format were collected retrospectively from the
hospital’s PACS. These scans then got  pseudonymized to protect patient's privacy. Next, the dataset was annotated by radiologists via a web-based labeling tool called VinDr Lab, which was developed to manage medical image labeling projects with dedicated features for the medical domain. Finally, the annotated exams were split into a training set of 4,000 exams and a test set of 1,000 exams.}
    \label{fig:overview}
\end{figure}

This study was approved by the Institutional Review Board of the HMUH and H108. All the personally identifiable information and protected health information of patients were removed. Additionally, this project did not affect clinical care at these two hospitals; hence patient consent was waived. The creation of the VinDr-Mammo dataset involves three stages: data acquisition, mammography reading, and data stratification. An overview of the data creation process is illustrated in Figure \ref{fig:overview}.  

\subsection*{Data acquisition}
In this step, 20,000 mammography images in DICOM format from 5,000 mammography examinations were randomly sampled from the pool of all mammography examinations taken between 2018 and 2020 via the Picture Archiving and Communication System (PACS) of Hanoi Medical University Hospital (HMUH -- \url{https://hmu.edu.vn/}) and Hospital 108 (H108 -- \url{https://www.benhvien108.vn/home.htm}). As the exams were randomly selected, the dataset includes both screening and diagnostic exams and represents the real distribution of patient cohorts in these hospitals. All images have the image presentation intent type of "FOR PRESENTATION" as those of for-processing were not stored by the hospitals. Images were acquired on equipments from 3 vendors, namely SIEMENS, IMS, and Planmed. All radiographers working at these hospitals were trained and certified by HMUH. To ensure patient privacy is protected, identifiable patient information in DICOM tags is fully removed via a Python script. Only necessary information used for loading and processing DICOM images and patient demographic information, i.e., age, is retained. Besides DICOM meta-data, associated information might appear in the images, such as laterality and view of the image and sometimes the patient's name. As this textual information usually appears in the corners of the image, we remove them by setting to black all pixels in a rectangle at each corner. The size of the rectangle is determined by visually inspecting a subset of the collected dataset. To validate the  pseudonymization stage, both DICOM metadata and image are manually reviewed by human readers.

\subsection*{Mammography reading}
This dataset aims to provide both the overall assessment of the breast and information of local-level findings, which are essential to developing CADx and CADe systems for breast cancer screening. To this end, the 5,000 sampled exams containing 20,000 images were re-read, as the associated radiology reports do not indicate the exact locations of the findings.

The reading results follow the schema and lexicon of the Breast Imaging Reporting and Data System~\cite{birads_2013}. At the breast level, the overall BI-RADS assessment categories and breast density level (also termed breast composition) are provided. There are seven BI-RADS assessment categories, namely BI-RADS 0 (need additional imaging or prior examinations), BI-RADS 1 (negative), BI-RADS 2 (benign), BI-RADS 3 (probably benign), BI-RADS 4 (Suspicious), BI-RADS 5 (highly suggestive of malignancy) and BI-RADS 6 (known biopsy-proven). Since the biopsy results are not available, there is no presence of BI-RADS 6 in the re-reading process. Regarding the breast density level, its four categories are A (almost entirely fatty), B (scattered areas of fibroglandular), C (heterogeneously dense), and D (extremely dense). For the mammography findings, the list of findings provided in this dataset includes the mass, calcification, asymmetries, architectural distortion, and other associated features, namely suspicious lymph node, skin thickening, skin retraction, and nipple retraction. Each finding is marked by a bounding box to localize the abnormal finding. In the given finding list, BI-RADS assessment is provided for mass, calcification, asymmetries, and architectural distortion. Since the purpose of this dataset is for breast cancer screening, benign findings, i.e., findings of BI-RADS 2, are not reported to reduce the annotating time. Only findings of BI-RADS categories greater than 2, which are not confident of benign or likely to be malignant, are marked. More details of the reading reports are provided in supplementary materials. Figure \ref{fig:sample_exam} illustrates a sample mammography exam with both finding annotations and breast-level assessments reported by radiologists.

\begin{figure}[h]
    \centering
    \begin{subfigure}[b]{0.22\textwidth}
        \centering
        \includegraphics[width=\linewidth]{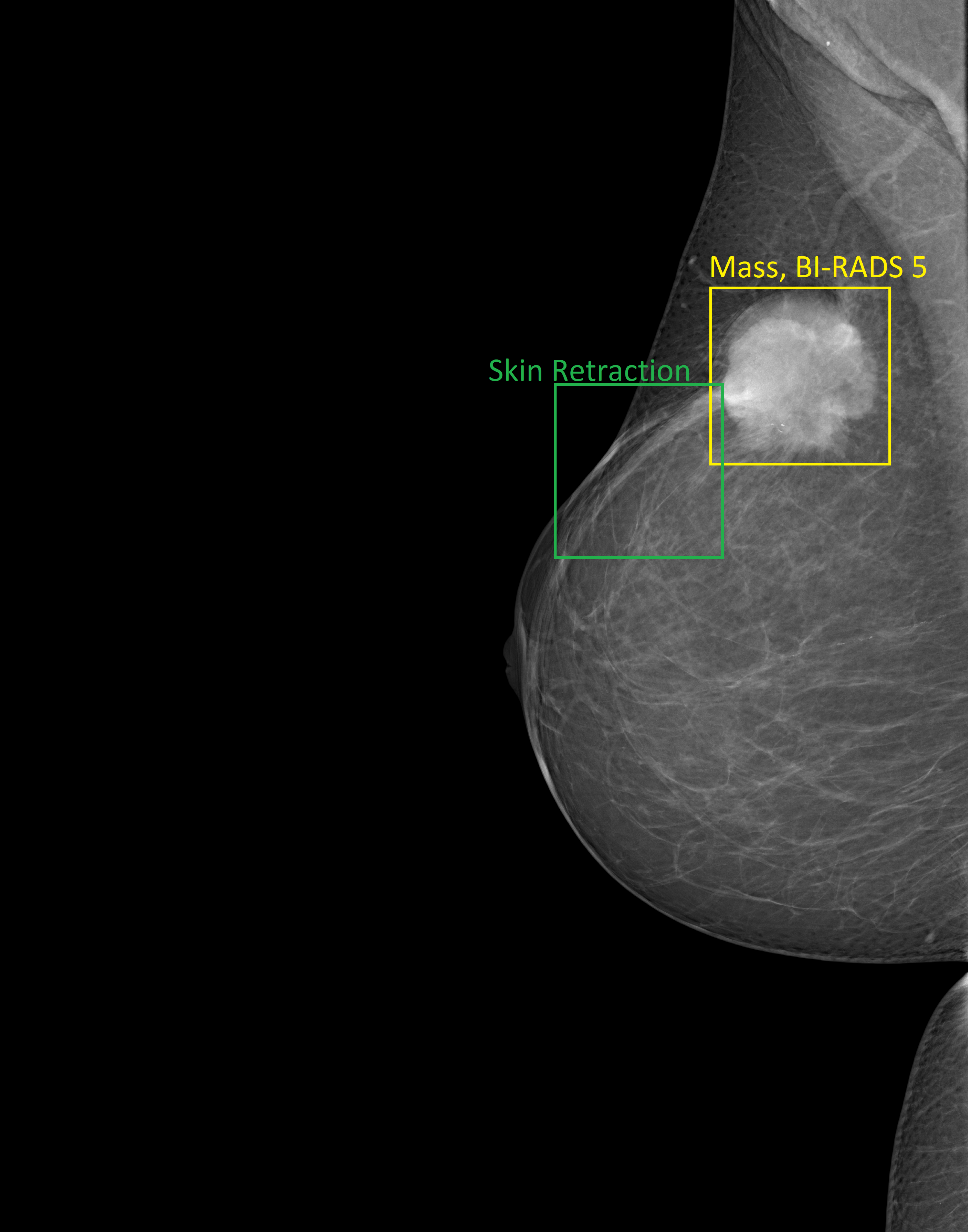}
        \caption{Right MLO}
    \end{subfigure} %
    \begin{subfigure}[b]{0.22\textwidth}
        \centering
        \includegraphics[width=\linewidth]{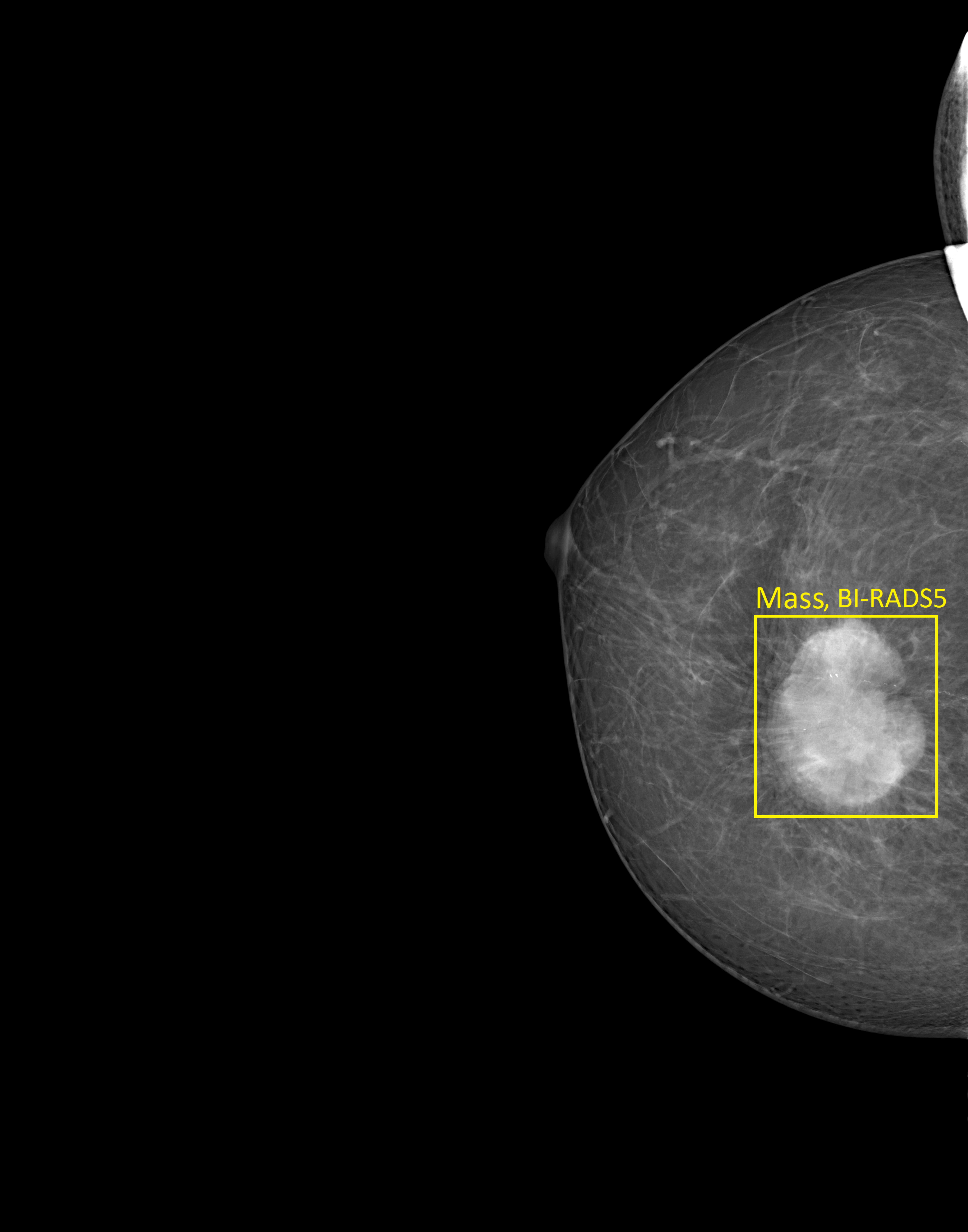}
        \caption{Right CC}
    \end{subfigure} %
    \begin{subfigure}[b]{0.22\textwidth}
        \centering
        \includegraphics[width=\linewidth]{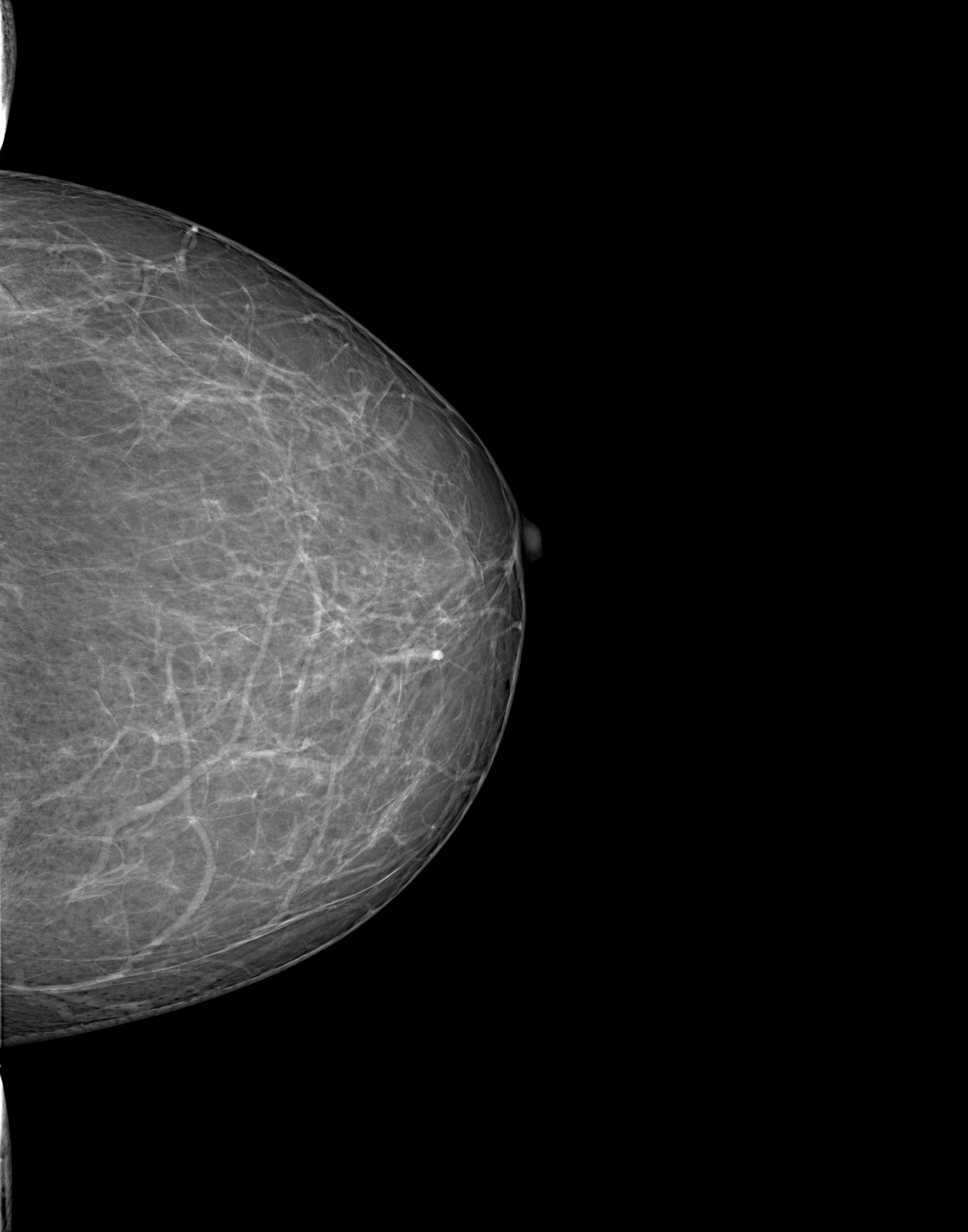}
        \caption{Left CC}
    \end{subfigure} %
    \begin{subfigure}[b]{0.22\textwidth}
        \centering
        \includegraphics[width=\linewidth]{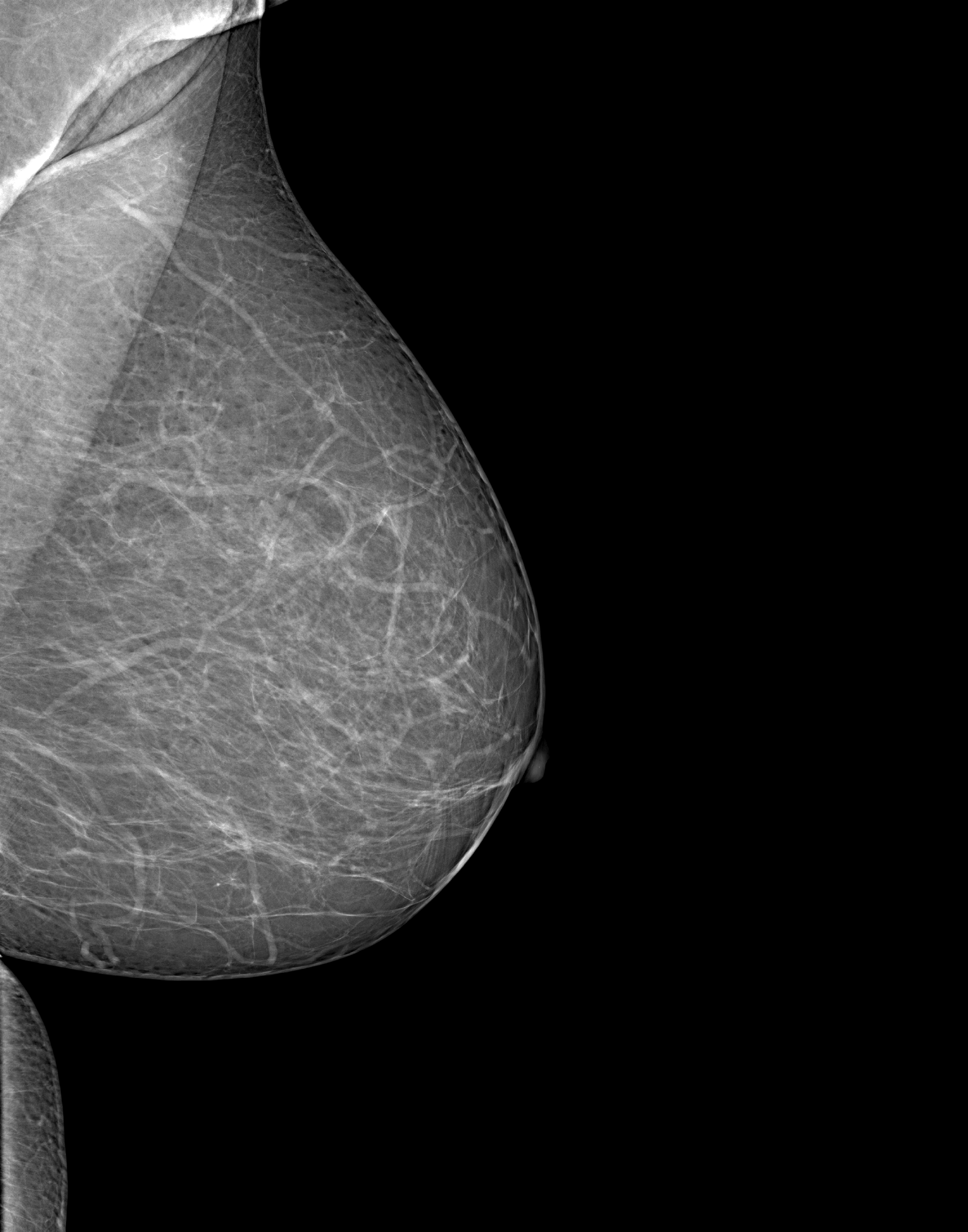}
        \caption{Left MLO}
    \end{subfigure} %

    \caption{A sample mammography exam with the right breast assessed with BI-RADS 5, density B and the left breast with BI-RADS 1, density B. CC denotes craniocaudal and MLO denotes mediolateral oblique.}
    \label{fig:sample_exam}
\end{figure}

The mammography reading process was facilitated by a web-based annotation tool called VinDr Lab (\url{https://github.com/vinbigdata-medical/vindr-lab}), which was specifically designed for viewing and annotating medical images with the medical image viewer being based on the Open Health Imaging Foundation project (\url{https://ohif.org/}). The three participating radiologists were able to remotely access the data for reading and annotating. All three radiologists hold healthcare professional certificates, which require up to eight years of the training program, and are approved by the Joint Review Committee on Education in Radiology, Vietnamese Ministry of Health. Furthermore, each reader has more than ten years of experience in the field. Specifically, all three radiologists received training in mammography interpretation and had an average of 19 years of clinical experience interpreting mammography (range 14–22 years). Additionally, each annotator reviewed an average of 13,333 mammography exams annually (range 10,000–15,000). Table \ref{rads} shows the characteristics of the radiologists who participated in our data annotation process.

\begin{table}[ht]
\centering 
\caption{\textsf{
Characteristics of the participating radiologists. Mean annual diagnostic volumes were estimated based on the number of mammogram scans interpreted. \label{rads}}}
\setlength{\tabcolsep}{5pt}
\renewcommand{\arraystretch}{1.5}
\begin{tabular}{P{80pt}P{80pt}P{160pt}}
\hline
\textbf{Annotator} & \textbf{Years' experience} & \textbf{Annual Diagnostic Volume (studies)} \\ \hline
Radiologist 1 & \hspace*{0.8cm} 14 & \hspace*{1.3cm} 10,000
\\ 
Radiologist 2 & \hspace*{0.8cm} 21 & \hspace*{1.3cm} 15,000
\\
Radiologist 3 & \hspace*{0.8cm} 22 & \hspace*{1.3cm} 15,000
\\ \hline
\textbf{Average} & \hspace*{0.8cm} \textbf{19} & \hspace*{1.3cm} \textbf{13,333}\\
\hline
\end{tabular}
\label{tab:existing-data}
\end{table}

Each mammography exam was then assigned to two radiologists and read independently. In cases of discordance, the exam would be assigned to the third radiologist at a higher senior experience level to make the final decision taking into account annotations of previous readers. After completing the reading process, the breast-level categories and local annotations were exported in JavaScipt Object Notation (JSON) format. Subsequently, we parsed the exported file to discard unnecessary information, namely annotation timestamp, and radiologist's identifier then simplified the file's structure and transformed it to comma-separated values (CSV) file so that it could be easily parsed.
\subsection*{Data stratification}

\begin{figure}[h]
    \centering
    \begin{subfigure}[b]{0.3\textwidth}
        \centering
        \includegraphics[width=\linewidth]{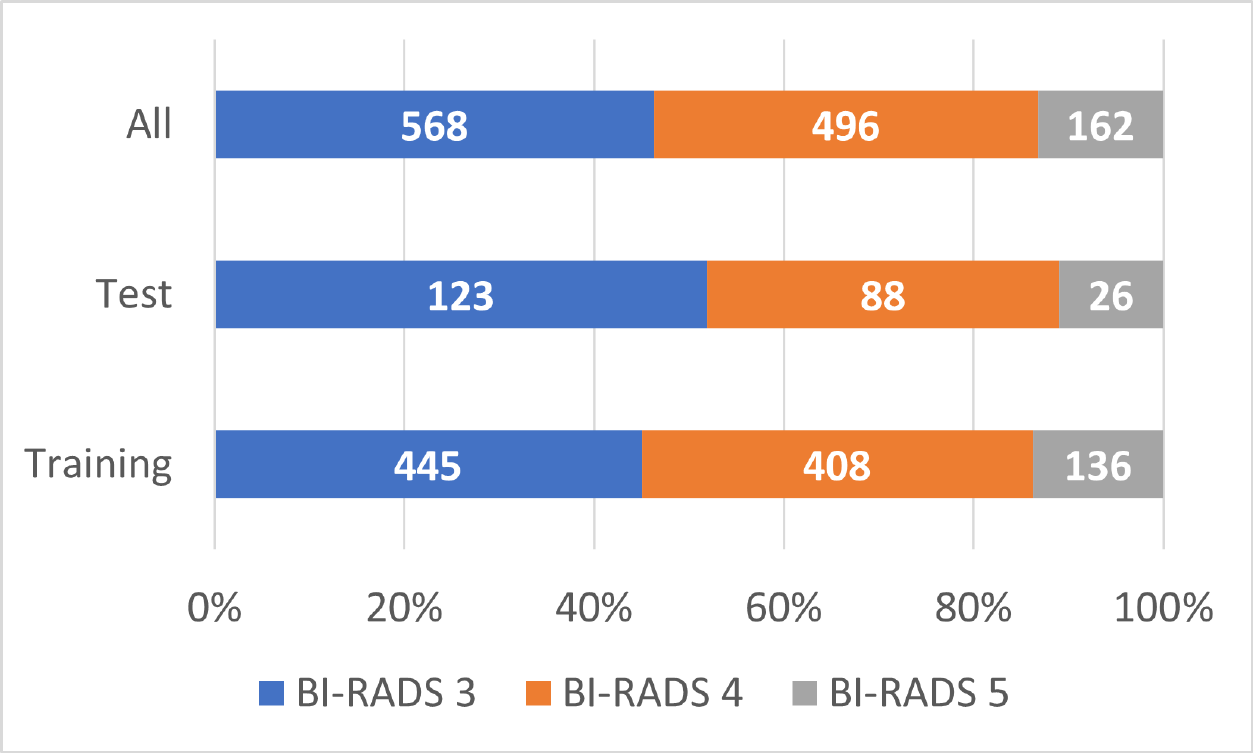}
        \caption{Mass}
    \end{subfigure} %
    \begin{subfigure}[b]{0.3\textwidth}
        \centering
        \includegraphics[width=\linewidth]{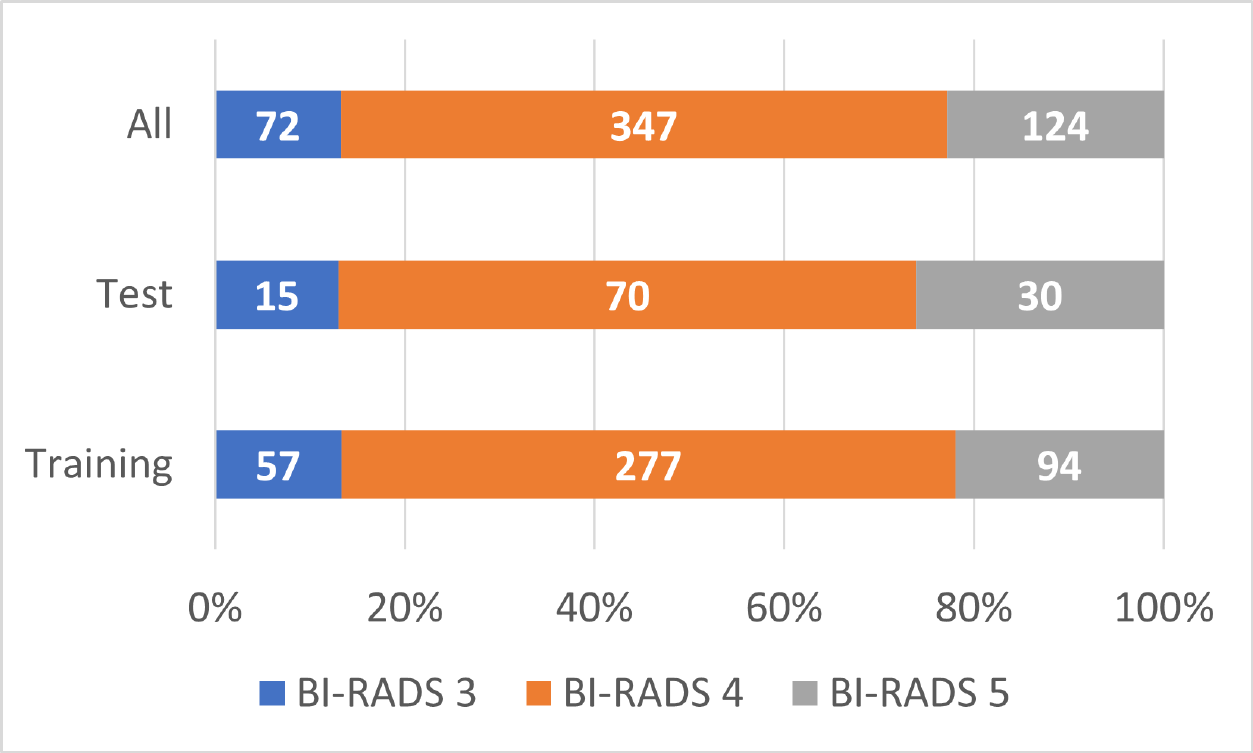}
        \caption{Suspicious Calcification}
    \end{subfigure} %
    \begin{subfigure}[b]{0.3\textwidth}
        \centering
        \includegraphics[width=\linewidth]{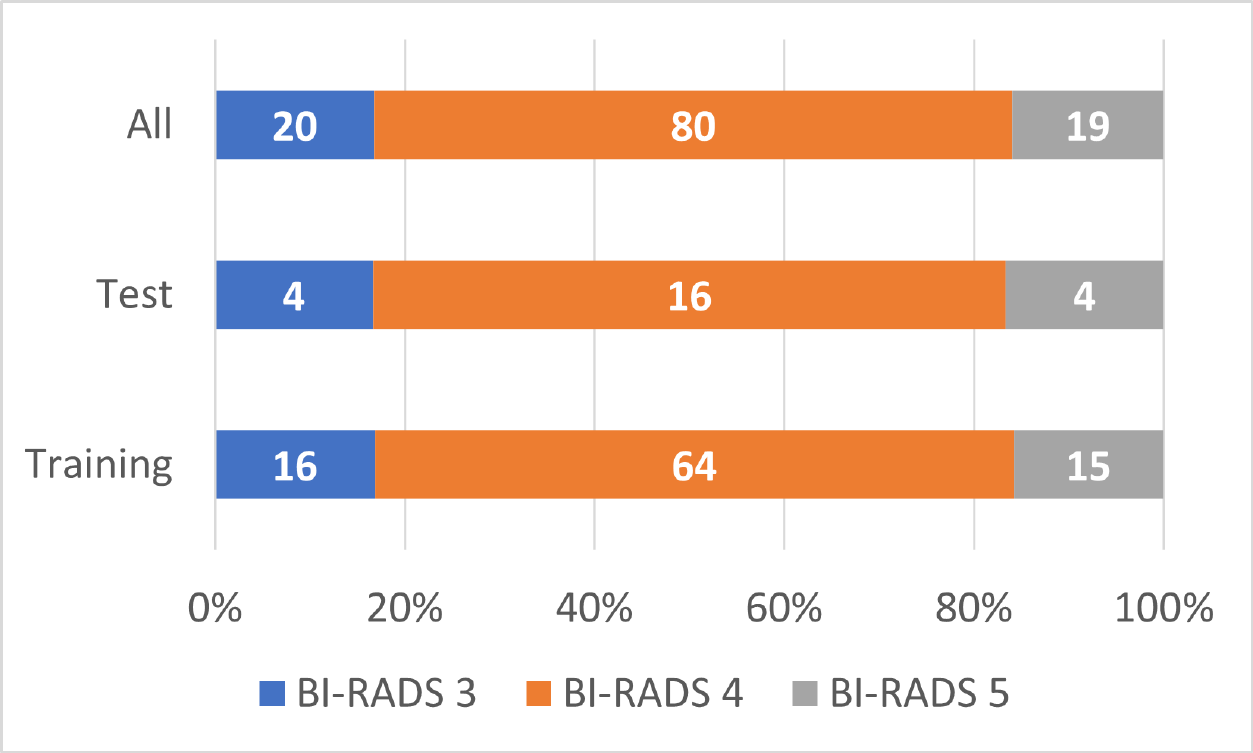}
        \caption{Architectural Distortion}
    \end{subfigure} %

    \begin{subfigure}[b]{0.3\textwidth}
        \centering
        \includegraphics[width=\linewidth]{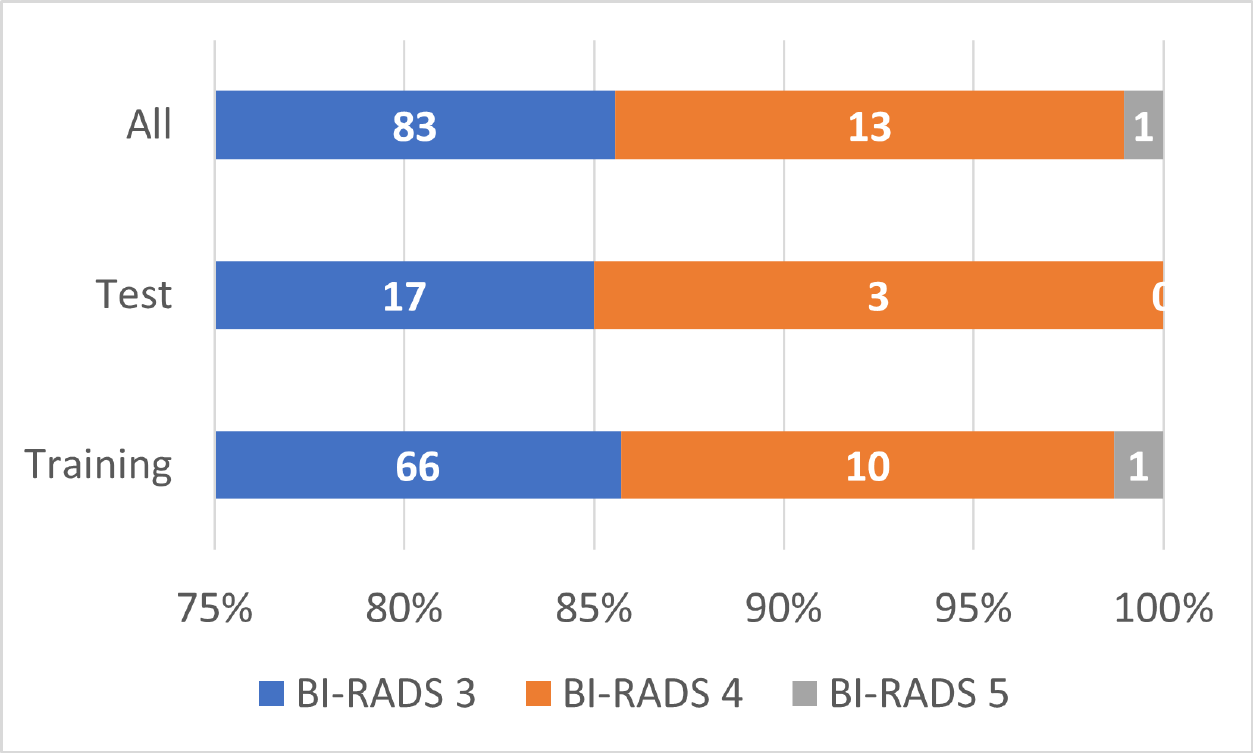}
        \caption{Asymmetry}
    \end{subfigure} %
    \begin{subfigure}[b]{0.3\textwidth}
        \centering
        \includegraphics[width=\linewidth]{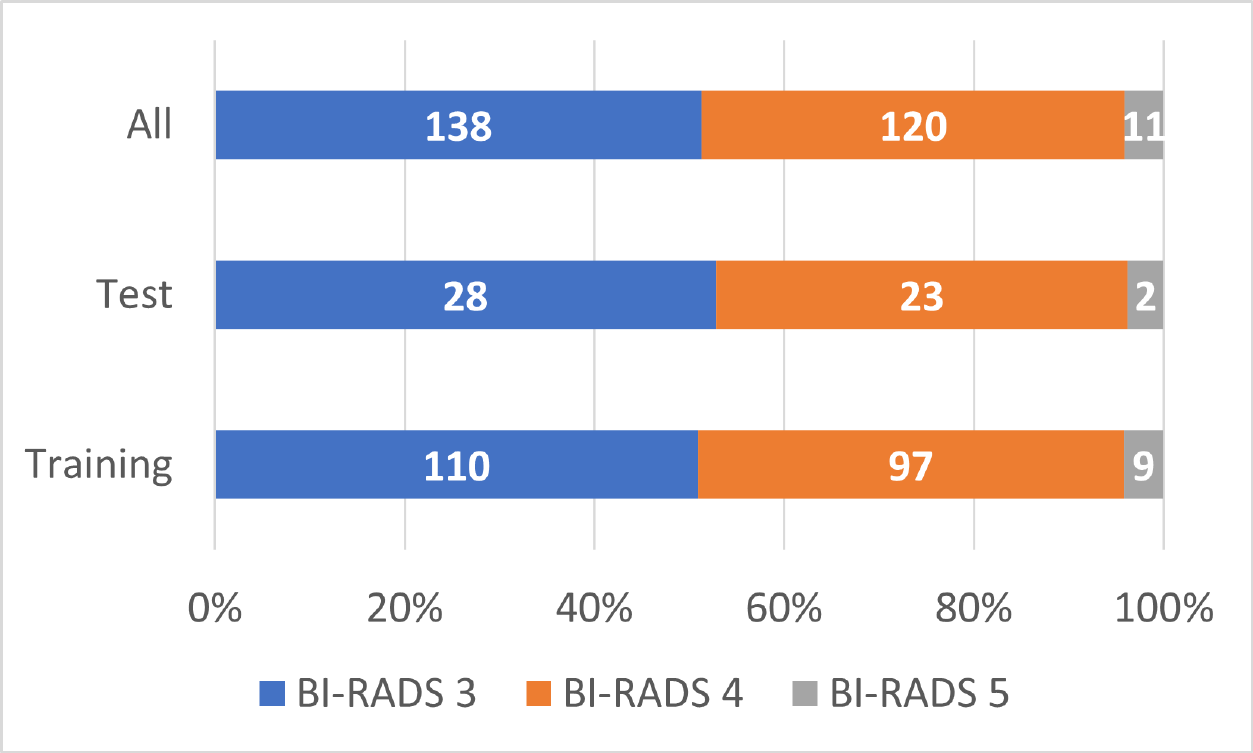}
        \caption{Focal Asymmetry}
    \end{subfigure} %
    \begin{subfigure}[b]{0.3\textwidth}
        \centering
        \includegraphics[width=\linewidth]{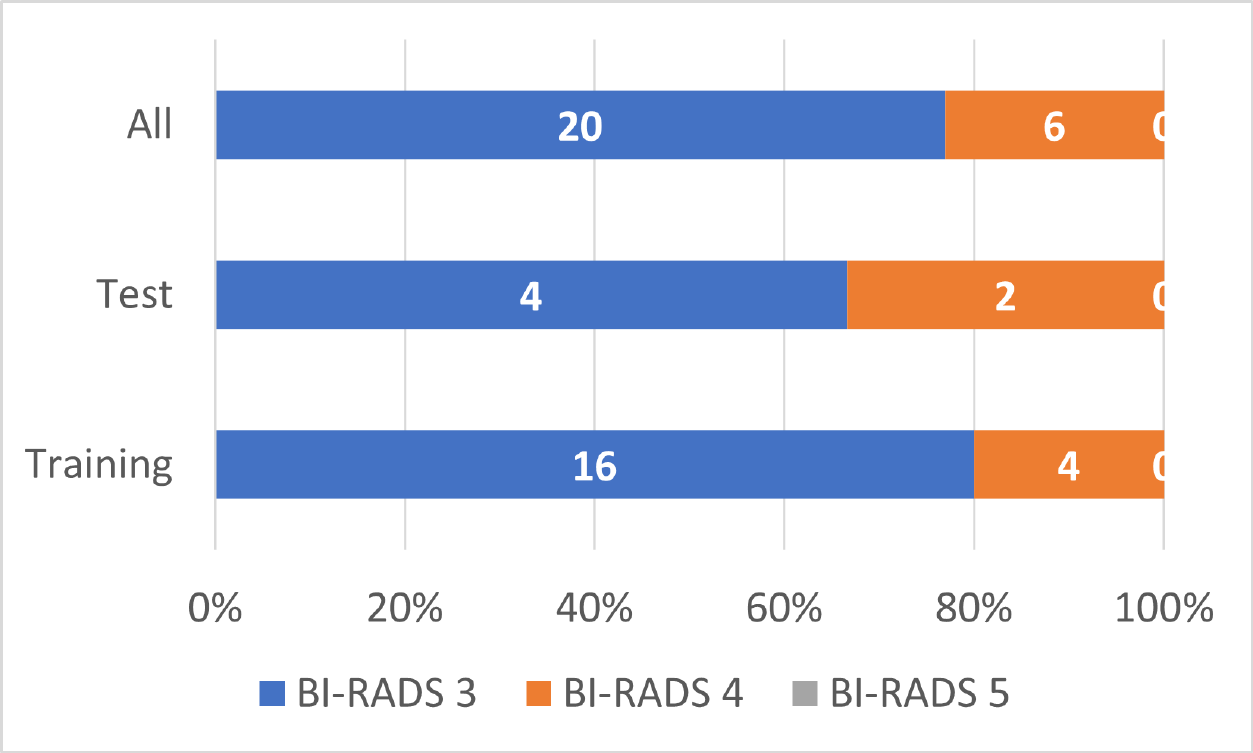}
      \caption{Global Asymmetry}
    \end{subfigure} %

    \caption{Statistics of BI-RADS assessment of findings.}
    \label{fig:finding_birads}
\end{figure}

\begin{figure}[h]
    \centering
    \includegraphics[width=0.6\linewidth]{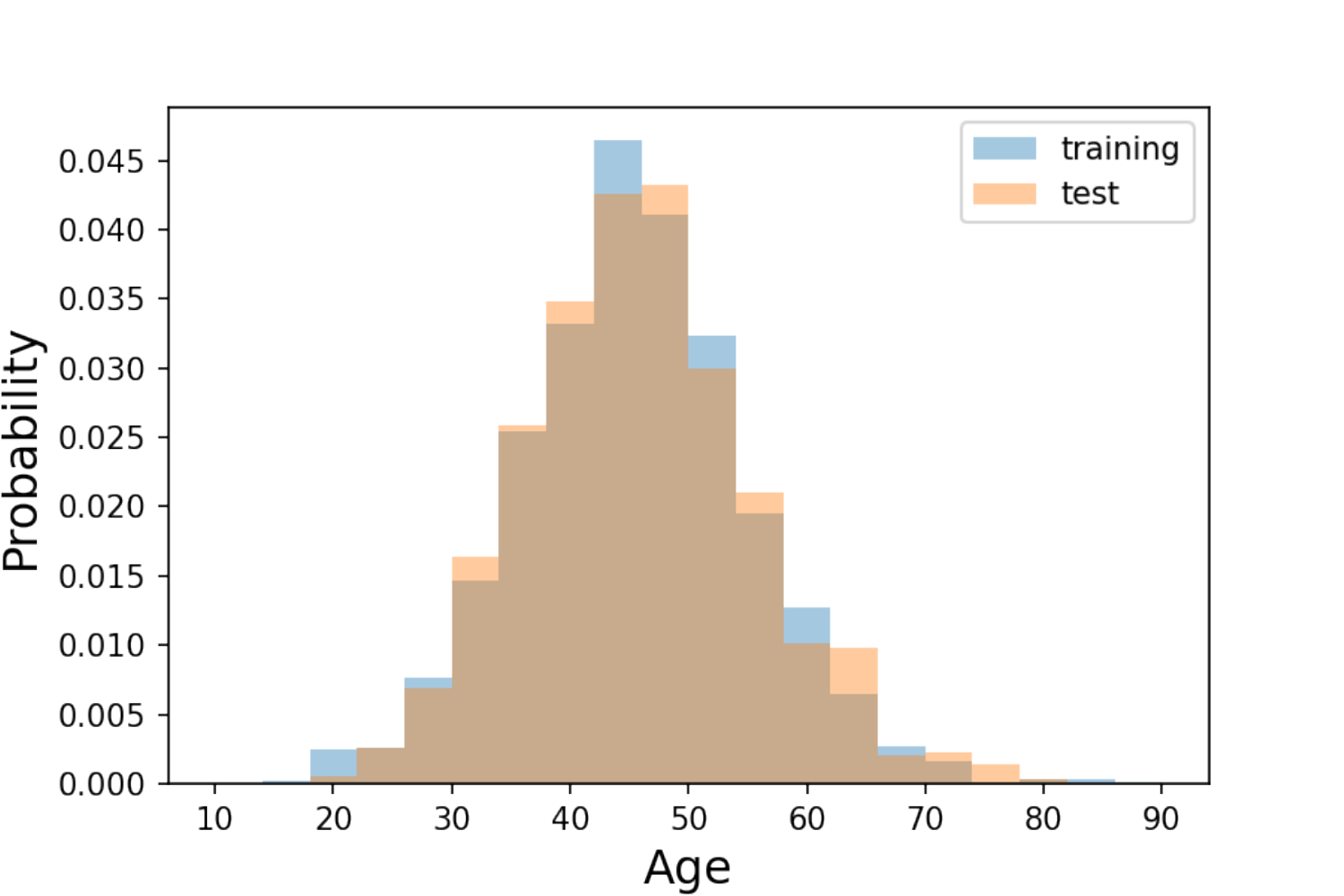}
    \caption{Distribution of patient age. This statistic is calculated overall all exams in which patient's age is available.}
    \label{fig:patient_age}
\end{figure}

Recent CADx and CADe solutions are mostly learning-based approaches that require separating the dataset into disjoint subsets for training and evaluation. A pre-defined training/test split would help guarantee that different research works will use the same exams for training and testing. Otherwise, inconsistent or unstated splits in different research works might hinder the reproducibility and comparison of these works. For an appropriate stratification, both the training and test sets should reflex the assessment, composition, and distribution of findings of the whole dataset. However, stratifying that dataset while preserving the correlation between various data characteristics is challenging as the number of combinations of different attributes grows exponentially with the number of attributes (in this case BI-RADS, breast composition, and findings categories). Hence, we split the dataset by an algorithm called iterative stratification~\cite{sechidis2011stratification} which bases on a relaxed target that only retains a fraction of the appearance of each attribute while ignoring their co-occurrence. One-fifth of the dataset, equivalent to 1,000 exams, is for testing and the rest for training. The attributes that are taken into account for splitting include breast-level BI-RADS categories, breast composition, findings categories, and the attached BI-RADS categories (if any). The distribution of breast-level BI-RADS categories, breast composition, and findings for each subset are provided in Table~\ref{tab:birads_dis}, Table~\ref{tab:density_dis}, and Table~\ref{tab:finding_dis}, respectively. The BI-RADS assessment of finding and patient age distribution are also depicted in Figure~\ref{fig:finding_birads} and Figure~\ref{fig:patient_age}.

\begin{table}[ht]
\centering
\setlength{\tabcolsep}{5pt}
\renewcommand{\arraystretch}{1.5}
\caption{\label{tab:birads_dis} Statistics of breast-level BI-RADS assessment.}
\begin{tabular}{rrrrrrr}
\hline
& \multicolumn{5}{c}{\textbf{Breast BI-RADS}} & \multicolumn{1}{l}{} \\ \cline{2-6}
& \multicolumn{1}{c}{\textbf{1}} & \multicolumn{1}{c}{\textbf{2}} & \multicolumn{1}{c}{\textbf{3}} & \multicolumn{1}{c}{\textbf{4}} & \multicolumn{1}{c}{\textbf{5}} & \multicolumn{1}{c}{\textbf{Total}} \\ \hline
\textbf{Training} & 5,362 (67.03\%) & 1,871 (23.39\%) & 372 (04.65\%) & 305 (03.81\%) & 90 (01.12\%) & 8,000 \\
\textbf{Test} & 1,341 (67.05\%) & 467 (23.35\%) & 93 (04.65\%) & 76 (03.80\%) & 23 (01.15\%) & 2,000 \\ \hline
\textbf{Overall} & 6,703 (67.03\%) & 2,338 (23.38\%) & 465 (04.65\%) & 381 (03.81\%) & 113 (01.13\%) & 10,000 \\ \hline
\end{tabular}

\end{table}

\begin{table}[ht]
\centering
\setlength{\tabcolsep}{5pt}
\renewcommand{\arraystretch}{1.5}
\caption{\label{tab:density_dis} Statistics of breast density.}
\begin{tabular}{rrrrrr}
\hline
& \multicolumn{4}{c}{\textbf{Breast Density}} & \\ \cline{2-5}
& \multicolumn{1}{c}{\textbf{A}} & \multicolumn{1}{c}{\textbf{B}} & \multicolumn{1}{c}{\textbf{C}} & \multicolumn{1}{c}{\textbf{D}} & \multicolumn{1}{c}{\textbf{Total}} \\ \hline
\textbf{Training} & 40 (00.50\%) & 764 (09.55\%) & 6,116 (76.45\%) & 1,080 (13.50\%) & 8,000 \\
\textbf{Test} & 10 (00.50\%) & 190 (09.50) & 1,530 (76.50\%) & 270 (13.50\%) & 2,000 \\ \hline
\textbf{Overall} & 50 (00.50\%) & 954 (09.54\%) & 7,646 (76.46\%) & 1,350 (13.50\%) & 10,000 \\ \hline
\end{tabular}
\end{table}

\begin{table}[ht]
\centering
\setlength{\tabcolsep}{3pt}
\renewcommand{\arraystretch}{1.5}
\caption{\label{tab:finding_dis}Findings statistics on the VinDr-Mammo dataset. The number of findings and the rate of findings per 100 images are provided for the training set, test set, and the whole dataset.}
\begin{tabular}{rrrr}
\hline
\multicolumn{1}{c}{\textbf{}} & \multicolumn{2}{c}{\textbf{Split}} & \\ \cline{2-3}
\multicolumn{1}{c}{\textbf{Finding}} & \multicolumn{1}{c}{\textbf{Training}} & \multicolumn{1}{c}{\textbf{Test}} & \multicolumn{1}{c}{\textbf{Total}} \\ \hline
Mass                        & 989 (6.181) & 237 (5.925) & 1,226 (6.130) \\
Suspicious Calcification    & 428 (2.675) & 115 (2.875) &  543 (2.715) \\
Asymmetry                   &  77 (0.481) &  20 (0.500) &   97 (0.485) \\
Focal Asymmetry             & 216 (1.350) &  53 (1.325) &  269 (1.345) \\
Global Asymmetry            &  20 (0.125) &   6 (0.150) &   26 (0.130) \\
Architectural Distortion    &  95 (0.594) &  24 (0.600) &  119 (0.595) \\
Skin Thickening             &  45 (0.281) &  12 (0.300) &   57 (0.285) \\
Skin Retraction             &  15 (0.094) &   3 (0.075) &   18 (0.090) \\
Nipple Retraction           &  30 (0.188) &   7 (0.175) &   37 (0.185) \\
Suspicious Lymph Node       &  46 (0.288) &  11 (0.275) &   57 (0.285) \\ \hline
\end{tabular}

\end{table}

\newpage

\section*{Data Records}
\label{sec:data_records}
Both DICOM images and radiologists' annotations of the dataset are available on PhysioNet~\cite{physionet} for public access. Breast-level and lesion-level annotations of the whole dataset are stored in CSV files \verb|breast-level_annotations.csv| and \verb|finding_annotations.csv|, respectively. The images are structured into subfolders according to the encoded study identifiers, each of which contains four images corresponding to four views of the exam. The subfolder name and image file name are named following the study identifier and image identifier. The information of the breast-level annotations is provided for each image even though there is redundancy since each breast is associated with two images of different view positions, i.e., MLO and CC. We find this representation more convenient because other metadata of the image, namely laterality and view position, can also be included, eliminating the need to parse this information from the DICOM tags. Metadata for each image in the \verb|breast-level_annotations.csv| file includes:
\begin{itemize}
\item \verb|study_id|: The encoded study identifier.
\item \verb|series_id|: The encoded series identifier.
\item \verb|image_id|: The encoded image identifier.
\item \verb|laterality|: Laterality of the breast depicted in the image. Either \verb|L| or \verb|R|.
\item \verb|view_position|: Breast projection. Standard views are CC and MLO.
\item \verb|height|: Height of the image.
\item \verb|width|: Width of the image.
\item \verb|breast_birads|: BI-RADS assessment of the breast that the image depicts.
\item \verb|breast_density|: Density category of the breast that the image depicts.
\item \verb|split|: Indicating the split to which the image belongs. Either \verb|training| or \verb|test|. 
\end{itemize}
Regarding breast findings, each annotation represents the occurrence of breast abnormality at a region, represented by a bounding box, in a specific image. This means that a single finding may associate with annotations from different views, yet this linking information is not acquired in the annotation process. Metadata for each finding annotation in the \verb|finding_annotations.csv| file contains:
\begin{itemize}
\item \verb|image_id|: The encoded identifier of the image in which the finding appears.
\item \verb|study_id|: The encoded identifier of the associated study.
\item \verb|series_id|: The encoded identifier of the associated series.
\item \verb|laterality|: Laterality of the breast in which the finding appears.
\item \verb|view_position|: Orientation with respect to the breast of the image.
\item \verb|height|: Height of the image.
\item \verb|width|: Width of the image.
\item \verb|breast_birads|: BI-RADS assessment of the breast that the image depicts.
\item \verb|breast_density|: Density category of the breast that the image depicts.
\item \verb|finding_categories|: List of finding categories attached to the region, e.g., mass with skin retraction.
\item \verb|finding_birads|: BI-RADS assessment of the marked finding.
\item \verb|xmin|: Left boundary of the box.
\item \verb|ymin|: Top boundary of the box.
\item \verb|xmax|: Right boundary of the box.
\item \verb|ymax|: Bottom boundary of the box.
\item \verb|split|: Indicating the split to which the image belongs. Either \verb|training| or \verb|test|. 
\end{itemize}

\section*{Technical Validation}

The data pseudonymization procedure and the quality of the labeling process were strictly controlled. First, all meta-data was manually reviewed to ensure that all individually identifiable health information or PHI~\cite{isola2019protected} of the patients has been fully removed to meet data privacy regulations such as the U.S. HIPAA~\cite{assistance2003summary} and the European GDPR~\cite{gdpr}. In addition, the image content of all mammograms was manually reviewed case-by-case by human readers to ensure that no patient information remained. We developed a set of rules underlying our labeling tool to reduce mislabeling. These rules allowed us to  verify the radiologist-generated labels automatically. Specifically, they prevent annotators from mechanical mistakes like forgetting to choose global labels or marking lesions on the image while choosing ``\verb|BI-RADS 1|'' as the breast-level assessment.

\section*{Usage Notes}

The VinDr-Mammo dataset was created for the purpose of developing and evaluating computer-aided detection and diagnosis algorithms based on full-field digital mammography. In addition, it can also be used for general tasks in computer vision, such as object detection and multiple-label image classification. To download and explore this dataset, users are required to accept a Date Usage Agreement (DUA) called PhysioNet Credentialed Health Data License 1.5.0 (\href{https://www.physionet.org/about/licenses/physionet-credentialed-health-data-license-150/}{https://www.physionet.org/about/licenses/physionet-credentialed-health-data-license-150/}). By accepting this DUA, users agree that the dataset can be used for scientific research and educational purposes only and will not attempt to re-identify any patients, institutions or hospitals. Additionally, the authors should cite this original paper for any publication that explores this dataset.

In this study, our objective is to provide an extensive open dataset of mammograms that include annotations from radiologists. We have used the consensus among radiologists as ground truth to ensure the reliability of the annotations. However, the VinDr-Mammo dataset has certain limitations such as the absence of pathology-confirmed ground truth data and other essential clinical information like molecular and histology data. As a result, it relies heavily on the expertise of radiologists. Biopsy tests are currently the most reliable means of measuring breast cancer. However, obtaining a significant number of mammographic images, each with a biopsy test, is impractical and outside the scope of this study. Given the data's incomplete support from pathology reports, it should not be used to directly evaluate CAD for diagnosis purposes, but only used in training settings. For the screening purpose, the dataset can be directly used to evaluate CAD after converting the provided BI-RADS annotations in 5 categories to the 3-category system: BI-RAS 0 (recall, correspond to \verb|BI-RADS 3|, \verb|BI-RADS 4|, \verb|BI-RADS 5|), BI-RAS 1 (normal, correspond to \verb|BI-RADS 1|), BI-RAS 2 (benign, correspond to \verb|BI-RADS 2|). Additionally, some abnormalities, such as skin and nipple retraction, have less than 40 samples, making studying these abnormalities on this dataset less reliable. Finally, the introduced dataset is not DICOM-compliant and it would fail to be processed properly by DICOM processing libraries.

\section*{Code Availability}
\label{sec:code}
The codes used in this study were made publicly available. The scripts used for loading and processing DICOM images are based on the following open-source repositories: Python 3.8.0 (\href{https://www.python.org/}{https://www.python.org/}); Pydicom 1.2.0 (\href{https://pydicom.github.io/}{https://pydicom.github.io/}); and Python hashlib (\href{https://docs.python.org/3/library/hashlib.html}{https://docs.python.org/3/library/hashlib.html}). The code for data pseudonymization and stratification was made publicly available at \href{https://github.com/vinbigdata-medical/vindr-mammo}{https://github.com/vinbigdata-medical/vindr-mammo}.

\bibliography{sample}

\section*{Acknowledgements}

We would like to acknowledge the Hanoi Medical University Hospital and Hospital 108 for their collaboration in creating the VinDr-Mammo dataset and for agreeing to make it publicly available. We are especially thankful to the radiologist team Nhung Hong Luu, Minh Thi Ngoc Nguyen, Huong Thu Lai, and other collaborators who participated in the data collection and labeling process.


\section*{Author contributions}

H.Q.N. and H.H.P designed the study; H.T.N performed the data pseudonymization and data stratification; H.H.P and H.T.N conducted the data acquisition and analysis; H.H.P. and H.T.N wrote the paper; all authors reviewed the manuscript.

\section*{Competing interests} 

This work was funded by the Vingroup JSC. The funder had no role in study design, data collection, and analysis, decision to publish, or preparation of the manuscript.


\newpage


\end{document}





\begin{center}
\textbf{\Large{Supplementary materials}}
\end{center} 

\begin{table*}[hbt!]
\caption{\textsf{Definition of findings used in the study.}}
\label{label-defination}
\centering
\setlength{\tabcolsep}{3pt}
\renewcommand{\arraystretch}{1.5}

\begin{tabular}{p{10pt}|p{110pt}|p{300pt}}
\hline
\multicolumn{2}{l|}{\textbf{Finding}} & \textbf{Definition}\\
\hline
\multicolumn{2}{l|}{1. Mass} & A mass is 3-dimensional and occupies space. It has completely or partially convex-outward borders and (when radiodense) appears denser in the center than at the periphery. If a potential mass is seen only on a single projection, it should be called an asymmetry until its 3-dimensionality is confirmed. \\
\hline
\multicolumn{2}{l|}{2. Suspicious Calcification} & Calcification with suspicious morphology (amorphous, coarse heterogeneous,  fine pleiomorphic, fine linear, or fine linear branching) or probably benign (BI-RADS 3).\\
\hline
\multirow{ 3}{*}{\rotatebox[origin=c]{90}{\parbox[c]{7cm}{\centering \textbf{Asymmetries}}}}
& 3. Asymmetry & This is an area of fibroglandular-density tissue that is visible on only one mammographic projection. Most such findings represent summation artifacts, a superimposition of normal breast structures, whereas those confirmed to be real lesions (by a subsequent demonstration on at least one more projection) may represent one of the other types of asymmetry or a mass.\\
\cline{2-3}
& 4.  Global Asymmetry & Global asymmetry is judged relative to the corresponding area in the contralateral breast and represents a large amount of fibroglandular-density tissue over a substantial portion of the breast (at least one quadrant). There is no mass, distorted architecture, or associated suspicious calcifications.\\
\cline{2-3}
& 5.  Focal Asymmetry &  A focal asymmetry is judged relative to the corresponding location in the contralateral breast, and represents a relatively small amount of fibroglandular-density tissue over a confined portion of the breast (less than one quadrant). It is visible on and has a similar shape on different mammographic projections (hence, a real finding rather than the superimposition of normal breast structures), but it lacks the convex-outward borders and the conspicuity of a mass. Rather, the borders of a focal asymmetry are concave-outward, and it is usually seen to be interspersed with fat. \\
\hline
\multicolumn{2}{l|}{6. Architecture Distortion} & The parenchyma is distorted with no definite mass visible. This includes thin straight lines or spiculations radiating from a point, and focal retraction,  distortion, or straightening at the anterior or posterior edge of the parenchyma.\\
\hline
\multicolumn{2}{l|}{7. Suspicious Lymph Node} & Axillary lymph nodes receive lymph from vessels that drain the arm, the walls of the thorax, the breast, and the upper walls of the abdomen. Features of suspicious lymph nodes include loss or disruption of central fatty hilum, loss or pericapsular fat line, irregular outer margins, hyperattenuating, and calcified.\\
\hline
\multicolumn{2}{l|}{10. Skin Thickening} & Skin thickening may be focal or diffuse and is defined as being greater than 2 mm in thickness. This finding is of particular concern if it represents a change from previous mammography examinations. However, unilateral skin thickening is an expected finding after radiation therapy.\\
\hline
\multicolumn{2}{l|}{11. Skin Retraction} & The skin is pulled in abnormally. \\
\hline
\multicolumn{2}{l|}{12. Nipple Retraction} &  The nipple is pulled in. This should not be confused with nipple inversion, which is often bilateral and which in the absence of any suspicious findings and when stable for a long period of time, is not a sign of malignancy. However, if nipple retraction is new, suspicion of underlying malignancy is increased. \\
 \hline 
\end{tabular}
\end{table*}

\begin{table}
\centering
\caption{\textbf{DICOM tags (a).} The list of DICOM tags that were retained for loading and processing raw images. All other tags were removed for protecting patient privacy. Details about all these tags can be found from DICOM Standard Browser at \href{https://dicom.innolitics.com/ciods}{https://dicom.innolitics.com/ciods}.}
\label{dicom_tags_retained}
\setlength{\tabcolsep}{3pt}
\renewcommand{\arraystretch}{1.5}
\begin{tabular}{p{60pt}|p{120pt}|p{300pt}}
\hline
\textbf{DICOM Tag} & 
\textbf{Attribute Name} & \textbf{Description} \\
\hline

(0010,0040) & Patient's Sex                      & Sex of named Patient.                                                                                                                                                                                                       \\
(0010,1010) & Patient's Age                      & Age of the Patient.                                                                                                                                                                                                         \\
(0010,1020) & Patient's Size                     & Length or size of the Patient, in meters.                                                                                                                                                                                   \\
(0010,1030) & Patient's Weight                   & Weight of the Patient, in kilograms                                                                                                                                                                                         \\
(0028,0010) & Rows                               & Number of rows in the image.                                                                                                                                                                                                \\
(0028,0011) & Columns                            & Number of columns in the image.                                                                                                                                                                                             \\
(0028,0030) & Pixel Spacing                      & Physical distance in the patient between the center of each pixel, specified by a numeric pair – adjacent row spacing (delimiter) adjacent column spacing, in mm.                                                           \\
(0018,1164) & Imager Pixel Spacing               & Physical distance measured at the front plane of the Image Receptor housing between the center of each pixel. Specified by a numeric pair – adjacent row spacing (delimiter) adjacent column spacing, in mm.                \\
(0028,0120) & Pixel Padding Value                & Single pixel value or one limit (inclusive) of a range of pixel values used in an image to pad to rectangular format or to signal background that may be suppressed.                                                        \\
(0028,0121) & Pixel Padding Range Limit          & Pixel value that represents one limit (inclusive) of a range of padding values used together with Pixel Padding Value (0028,0120) as defined above.                                                                         \\
(0028,0100) & Bits Allocated                     & Number of bits allocated for each pixel sample. Each sample shall have the same number of bits allocated. Bits Allocated (0028,0100) shall be either 1, or a multiple of 8.                                                \\
(0028,0101) & Bits Stored                        & Number of bits stored for each pixel sample. Each sample shall have the same number of bits stored.                                                                                                                         \\
(0028,0102) & High Bit                           & Most significant bit for pixel sample data. Each sample shall have the same high bit. High Bit (0028,0102) shall be one less than Bits Stored (0028,0101)                                                                   \\
(0028,0103) & Pixel Representation               & Data representation of the pixel samples. Each sample shall have the same pixel representation.                                                                                                                             \\
(2050,0020) & Presentation LUT Shape            & Specified predefined Presentation LUT transformation. Required of Presentation LUT Sequence (2050,0010) is absent.                                                                                                          \\
(0028,0106) & Smallest Image Pixel Value         & The minimum actual pixel value encountered in this image.                                                                                                                                                                   \\
(0028,0107) & Largest Image Pixel Value          & The maximum actual pixel value encountered in this image.                                                                                                                                                                   \\
(0028,1050) & Window Center                      & Window Center for display.                                                                                                                                                                                                  \\
(0028,1051) & Window Width                       & Window Width for display.                                                                                                                                                                                                   \\
(0028,1055) & Window Center \& Width Explanation & Free form explanation of the meaning of the Window Center and Width. Multiple values corresponding to multiple Window Center and Width values.                                                                              \\
(7FE0,0010) & Pixel Data                         & A data stream of the pixel samples that comprise the Image. Required if Pixel Data Provider URL (0028,7FE0) is not present.                                                                                                 \\

\hline
\end{tabular}
\end{table}

\begin{table}
\centering
\caption{\textbf{DICOM tags (b).} The list of DICOM tags that were retained for loading and processing raw images. All other tags were removed to protect patient privacy. Details about all these tags can be found from DICOM Standard Browser at \href{https://dicom.innolitics.com/ciods}{https://dicom.innolitics.com/ciods}.}
\label{dicom_tags_retained}
\setlength{\tabcolsep}{3pt}
\renewcommand{\arraystretch}{1.5}
\begin{tabular}{p{60pt}|p{120pt}|p{300pt}}
\hline
\textbf{DICOM Tag} & 
\textbf{Attribute Name} & \textbf{Description} \\
\hline

(0028,1056) & VOI LUT Function                   & Describe a VOI LUT function to apply to the values of Window Center (0028,1050) and Window Width (0028,1051).                                                                                                               \\
(0028,3010) & VOI LUT Sequence                   & Defines a Sequence of VOI LUTs. One or more items shall be included in this Sequence. The required of Window Center (0028,1050) is not present. May be present otherwise.                                                      \\
(0028,3002) & LUT Descriptor                     & Specifies the format of the LUT Data in this Sequence.                                                                                                                                                                      \\
(0028,3003) & LUT Explanation                    & Free form text explanation of the meaning of the LUT.                                                                                                                                                                       \\
(0028,3006) & LUT Data                           & LUT Data in this sequence.                                                                                                                                                                                                  \\
(0028,1052) & Rescale Intercept                  & The value b in the relationship between stored values (SV) and the output units. Output units = m*SV+b                                                                                                                          \\
(0028,1053) & Rescale Slope                      & m in the equation specified in Rescale Intercept (0028,1052).                                                                                                                                                               \\
(0028,1054) & Rescale Type                       & Specifies the output units of Rescale Slope (0028,1053) and Rescale Intercept(0028,1052).                                                                                                                                   \\
(0028,0004) & Photometric Interpretation         & Specifies the intended interpretation of the pixel data.                                                                                                                                                                    \\
(0028,2110) & Lossy Image Compression            & Specifies whether an Image has undergone lossy compression (at a point in its lifetime).                                                                                                                                   \\
(0028,2112) & Lossy Image Compression Ratio      & Describes the approximate lossy compression ratio(s) that have been applied to this image.                                                                                                                                  \\
(0028,2114) & Lossy Image Compression Method     & A label for the lossy compression method(s) that have been applied to this image.                                                                                                                                           \\
(0028,0002) & Samples per Pixel                  & Number of samples (planes) in this image.                                                                                                                                                                                   \\
(0028,0008) & Number of Frames                   & Number of frames in a Multi-frame Image.                                                                                                                                                                                    \\
(0008,0018) & SOP Instance UID                   & Uniquely identifies the SOP Instance                                                                                                                                                                                        \\
(0020,000e) & Series Instance UID                & Unique identifier of the Series containing the referenced Instances.                                                                                                                                                        \\
(0020,000d) & Study Instance UID                 & Unique identifier of the Study containing the referenced Instances.                                                                                                                                                         \\
(0008,0060) & Modality                           & Type of equipment that originally acquired the data used to create the images in this Series.                                                                                                                                \\
(0018,0015) & Body Part Examined                 & Text description of the part of the body examined.                                                                                                                                                                          \\
(0008,0068) & Presentation Intent Type           & Identifies the intent of the images that are contained within this Series                                                                                                                                                   \\
(0008,0070) & Manufacturer                       & Manufacturer of the device.                                                                                                                                                                                                 \\
(0008,1090) & Manufacturer’s Model Name          & Manufacturer’s model name of the device.                                                                                                                                                                                    \\
(0020,0060) & Laterality                         & Laterality of (paired) body part examined. Required if the body part examined is a paired structure and Image Laterality (0020,0062) or Frame Laterality (0020,9072) or Measurement Laterality (0024,0113) are not present. \\
(0020,0062) & Image Laterality                   & Laterality of (possibly paired) body part (as described in Anatomic Region Sequence (0008,2218)) examined.                                                                                                                  \\
(0018,0051) & View Position                      & Radiographic view of the image relative to the imaging subject’s orientation.
\\

\hline
\end{tabular}
\end{table}
